\documentclass[11pt]{article}
\usepackage{moriond,epsfig}
\usepackage{amsmath}

\def\Journal#1#2#3#4{{#1} {\bf #2}, #3 (#4)}

\def\PRL{\em Phys. Rev. Lett.}

\begin{document}
\vspace*{4cm}

\title{FORWARD PHYSICS AT THE LHC}

\author{XAVIER ROUBY,\\on behalf of CMS and ATLAS Collaborations}

\address{Universit\'e catholique de Louvain,\\Center for Particle Physics and Phenomenology,\\Chemin du Cyclotron 2,\\1348 Louvain-la-Neuve, Belgium}

\maketitle\abstracts{
The CMS/TOTEM and ATLAS collaborations carry out a program of forward physics with several near-beam detectors extending their coverage to high pseudorapidities. This instrumentation includes calorimeters (CASTOR and ZDC), tracking devices (TOTEM T1 and T2), proton taggers (Roman Pots and FP420), and instrumentation dedicated to luminosity monitoring and normalisation. A rich physics program is accessible, including soft QCD interactions, Diffraction, photon-induced physics and luminosity measurements.
}

\section{Introduction}
The Large Hadron Collider (LHC) is a powerful collider, located at \textsc{cern} (Geneva). It will accelerate two proton beams, and collide them at $14~\textrm{TeV}$, reaching unprecedented centre-of-mass energies. The LHC intensity of collisions, or \textit{luminosity} $\mathcal{L}$, will grow from about $10^{27}$ to $10^{34}~\textrm{cm}^{-2}\textrm{s}^{-1}$. Heavy Ion collisions are also foreseen~\cite{lhc}, but these are not discussed in this document. The total LHC cross section is predicted by the COMPETE collaboration~\cite{co}:
\begin{equation} \sigma_\textrm{tot}(LHC) = 111.5 \pm 1.2 (stat) ^{+4.1}_{-2.1} (syst)~\textrm{mb} \end{equation}

Half of this large value corresponds to very soft physics, including elastic $pp$ collisions. Five experiments, ATLAS, ALICE, CMS/TOTEM and LHCb observe collision final states at the four interaction regions, IR1, IR2, IR5 and IR8, respectively. 

ATLAS and CMS are general purpose experiments with central detectors 
with full $\phi$ coverage and a pseudorapidity~\footnote{The pseudorapidity $\eta$ is defined in terms of the polar angle $\theta$: $\eta = -\ln ( \tan \theta/2 )$.} $\eta$ coverage reaching $2.5$ for tracking systems and $5.0$ 
for calorimetry. The TOTEM experiment is an approved experiment for measuring $\sigma_\textrm{tot}$ and $\sigma_\textrm{elastic}$ and shares the interaction point with CMS. CMS and TOTEM plan to carry out a joint program on forward and diffractive physics\cite{opus} during LHC running with 
nominal beam optics.

The CMS and ATLAS detectors see most of the charged particles from collision final states, but still a significant amount of them escapes from detection, with an pseudorapidity $|\eta|$ above $5$. Moreover, most of the energy flow of final states is in very forward regions, well beyond their reach. Extra instrumentation located along the LHC beamline in forward regions extend the central detector acceptances, and are detailed in this document.


\section{Instrumentation around IP5\label{sec:ip5} and forward physics}
The CMS detector~\cite{cms} consists of a central tracker ($|\eta| \leq 2.5$), an electromagnetic ($|\eta| \leq 3.0)$ and a hadronic ($|\eta| \leq 5.0)$ calorimeters, and a muon system ($|\eta| \leq 2.4)$, all embedded in a solenoidal magnetic field peaking at $4~\textrm{T}$. It is completed on each side by the Zero Degree Calorimeters for the measurement of forward neutrals ($\gamma$ and $n$), with $|\eta| > 8.1$. The ZDCs are quartz-tungsten (QW) sampling calorimeters, with electromagnetic (EM) and hadronic (HAD) sections, and are located at $140~\textrm{m}$ away from the interaction point (inside the TAN neutral absorber, where the LHC beampipes separate). The CASTOR calorimeters will be installed~\footnote{One CASTOR will be installed in 2008, a second one in 2009 if funded.} to complement CMS calorimetry from $5.1$ to $6.6$ in pseudorapidity. They will also be QW sampling calorimeters, with EM and HAD sections. Finally, the possible FP420 detectors at $420~\textrm{m}$, for proton tagging, are discussed separately in section \ref{sec:fp420}.

The TOTEM experiment~\cite{totem} has a complementary $|\eta|$ coverage. TOTEM T1 is made of several Cathode Strip Chambers, providing tracking data for $3.1 \leq |\eta| \leq 4.7$, in front of the forward hadronic calorimeters of CMS. Similarly, TOTEM T2 provides tracking data in front of CASTOR calorimeters ($5.3 \leq |\eta| \leq 6.7$), with GEM detectors. At last, TOTEM roman pots RP220, located at $220~\textrm{m}$ sense with dedicated silicon microstrip detectors the forward proton scattered at IP, with some energy loss $E_\textrm{loss}$ such that $0.02 \leq \xi \leq 0.2$, with $\xi =E_\textrm{loss} / E_\textrm{beam}$.

With this forward instrumentation, a diverse physics program can be 
carried out:
\begin{itemize}
 \item \textbf{Diffraction}: 
 Both soft and hard diffraction can be studied. In the absence of
pile-up, diffractive processes can be selected by detecting the large
rapidity gap in their hadronic final state. This can be done by means of
vetoing activity in T1/HF, T2/Castor and ZDC. In the presence of pile-up,
a selection is possible by means of tagging the proton that escapes the
interaction intact. Diffraction with a hard scale can be studied in
diffractive production of, for example, $W$~\cite{SDW}, jets, heavy flavors,
following the measurements done at Tevatron. These include determination
of the rapidity gap survival probability and measuring diffractive PDFs.

 \item \textbf{High energy photon-induced interactions}, with photon-photon, photon-parton and photon-proton collisions. These interactions are very well known theoretically and yield to very clean final states. This includes the exclusive production of charged particles within and beyond the Standard Model (SM), for the luminosity measurement ($\mu^+ \mu^-$), quartic gauge couplings ($W^+ W^-$), Super-Symmetric direct observations (charged fermions and scalars). Photon-proton interactions includes the associated photoproductions of $W$ and $t$, $W$ and $H$, or anomalous production of single top via Flavour Changing Neutral Currents.

   \item \textbf{Low-$x$ dynamics}, with forward jet studies, and Drell-Yan events, allowing to constrain the proton PDF at low $x$ ($\mathcal{O}(10^{-6})$), which is a real extension of the current measurements. These measurements with the forward calorimeters will give data for the study of parton saturation, BFKL/CCFM dynamics, multi-parton scattering, underlying events and multiple interactions.

   \item \textbf{Large rapidity gaps} (LRGs) between forward jets. HERA and Tevatron observed events with hard scale and a large separation in pseudorapitidy between jets. The wide $|\eta|$ coverage is suitable for studying such events at IP5.

   \item \textbf{Cosmic ray physics}. The LHC $pp$ c.m.s. energy at 14 TeV corresponds to a fixed target energy of $100~\textrm{PeV}$. This will provide data for the validation of the showering models for the description of interactions between Ultra High Energy cosmic rays and Earth upper atmosphere. Forward calorimeters (HF, CASTOR, ZDC) will measure the energy and particle flows of interest.

   \item \textbf{Luminosity} monitoring or normalisation. This LHC parameter is important for the cross-section measurement of any process. 
An absolute measurement of the luminosity can be carried out by TOTEM
in runs with special LHC optics, and by CMS in nominal optics runs by
means of exclusive dimuon production. Luminosity monitoring will be done
in CMS with the help of the HF calorimeter and a dedicated Pixel
Luminosity Telescope\cite{cms}.
\end{itemize}

\section{Instrumentation around IP1\label{sec:ip1} and forward physics}
Similarly to CMS, the ATLAS experiment~\cite{atlas} is made of a central tracking system ($|\eta| \leq 2.5$), EM ($|\eta| \leq 3.2$) and HAD ($|\eta| \leq 4.9$) calorimeters and muon spectrometers ($|\eta| \leq 2.7$). The magnetic field is provided by a central solenoid (peaking at $2.6~\textrm{T}$), surrounded by air-core toroids arranged in an $8-\textrm{fold}$ geometry (up to $4.1~\textrm{T}$). Several instruments will complement ATLAS coverage:

\begin{itemize}
 \item LUCID detectors~\cite{lucid} are meant for relative luminosity measurement, using Cerenkov Imaging Detector. These are dedicated luminosity monitors with $20$ counters, covering $5.6 < |\eta| < 6.0$, and located at $17~\textrm{m}$ from IP1. They count tracks from minbias, diffractive events in a restricted phase space region. LUCID measurement relies on zero counting (low luminosities) or particle counting (any luminosity).
 \item Zero Degree Calorimeters (ZDCs)~\cite{atlaszdc}, at $140~\textrm{m}$ from IP1, cover the $|\eta| \geq 8.3$ region for neutrals. These are sampling calorimeters made of quartz fibres and W plates, and have both EM and HAD sections. Working in both $pp$ and $AA$ collisions, they are able to reconstruct $\pi^0$, $\eta$, $\eta^\prime$, $\Delta$, $\Sigma$ and $\Lambda$ from $n$ and $\gamma$ measurements. ZDCs serve also for Van der Meer scans or for counting neutrons in AA collisions, as do the CMS ZDCs.
 \item ALFA (Absolute Luminosity For ATLAS)~\cite{lucid} detectors are roman pot, located at $237~\textrm{m}$ from IP1 and based on scintillating fibres. Their tracking data will measure~\footnote{ALFA detectors are expected to be ready in 2009.} elastic scattering parameters, for luminosity normalisation (goal is about $3\%$), calibration of LUCID, measurement of the total cross section and studies of hard diffraction (proton tagging) in nominal optics runs in conjunction with main ATLAS detector. Luminosity can be measured via Coulomb scattering or the optical theorem (Eq.~\ref{eq:optical}):
\end{itemize}

The event rate $dN/dt$ is related to the luminosity $L$, the QED coupling constant $\alpha$, and the total cross section $\sigma_\textrm{tot}$. The $\rho$ factor is the ratio between real and imaginary parts of the elastic scattering amplitude at $\theta=0$. The $b$ parameter describes the evolution of the strong interaction amplitude with respect to the transfer momentum $|t|$. 
\begin{equation} 
\begin{split}
\textrm{Elastic scattering} ~~~ &\frac{dN}{dt} \approx L \pi \Big | - \frac{2 \alpha}{|t|} + \frac{\sigma_\textrm{tot}}{4 \pi} (i+ \rho) \exp(-b|t|/2) \Big |^2 \\
\textrm{Optical theorem} ~~~ &\frac{1}{L} = \frac{1}{16 \pi} \frac{\sigma^2_\textrm{tot} (1+\rho^2)}{\big |dN_\textrm{el}/dt \big | _{t=0} } \\
\end{split}
\label{eq:optical} 
\end{equation}

In addition, possible options are investigated by the collaboration: the RP220 project, for roman pots located at $220~\textrm{m}$ from IP1, for a forward proton tagging if $0.02 \leq \xi \leq 0.2$ and FP420 detectors at $420~\textrm{m}$ with $p-\textrm{tag}$ in $0.002 \leq \xi \leq 0.02$  (Sec.~\ref{sec:fp420}). 
ATLAS will use forward detectors to measure the total and elastic cross
section and carry out an absolute luminosity measurement. ATLAS also plans
to implement a program on soft and hard diffraction.

\section{FP420\label{sec:fp420}}
FP420 is an R\&D program~\cite{fp420}  currently under review in ATLAS and CMS. FP420
proposes the installation of fast timing detectors and forward proton taggers at $420~\textrm{m}$ from IP1 and IP5. 
These detector station will contain Si 3D sensors for the measurement of $p$ position and angle, and Cerenkov timing detectors (Quartic and GasTOF). Detectors will be attached to a moving beam pipe for approaching the beam for operation. 

FP420 program probes discovery physics via for instance Central Exclusive Productions (CEP) of Higgs or particles beyond the Standard Model, with tagged leading protons (e.g. $pp \rightarrow p \oplus H \oplus p$). The central system mass $M$ can be measured from the forward proton detection only ($\xi_1$ and $\xi_2$), using the Missing Mass method: $M^2 = \xi_1 \xi_2 s$ for a given c.m.s collision energy $s$.
For example, Higgs CEP cross sections range from $3~\textrm{fb}$ in SM to $10-100~\textrm{fb}$ in MSSM. 

Proton position and angle measurements are used for the reconstructed energy and momentum loss at IP. Timing data constraint the $z$ position of the event vertex, used for event pile-up rejection.

\section{Early forward physics example}
\begin{figure}
\centering 
\psfig{figure=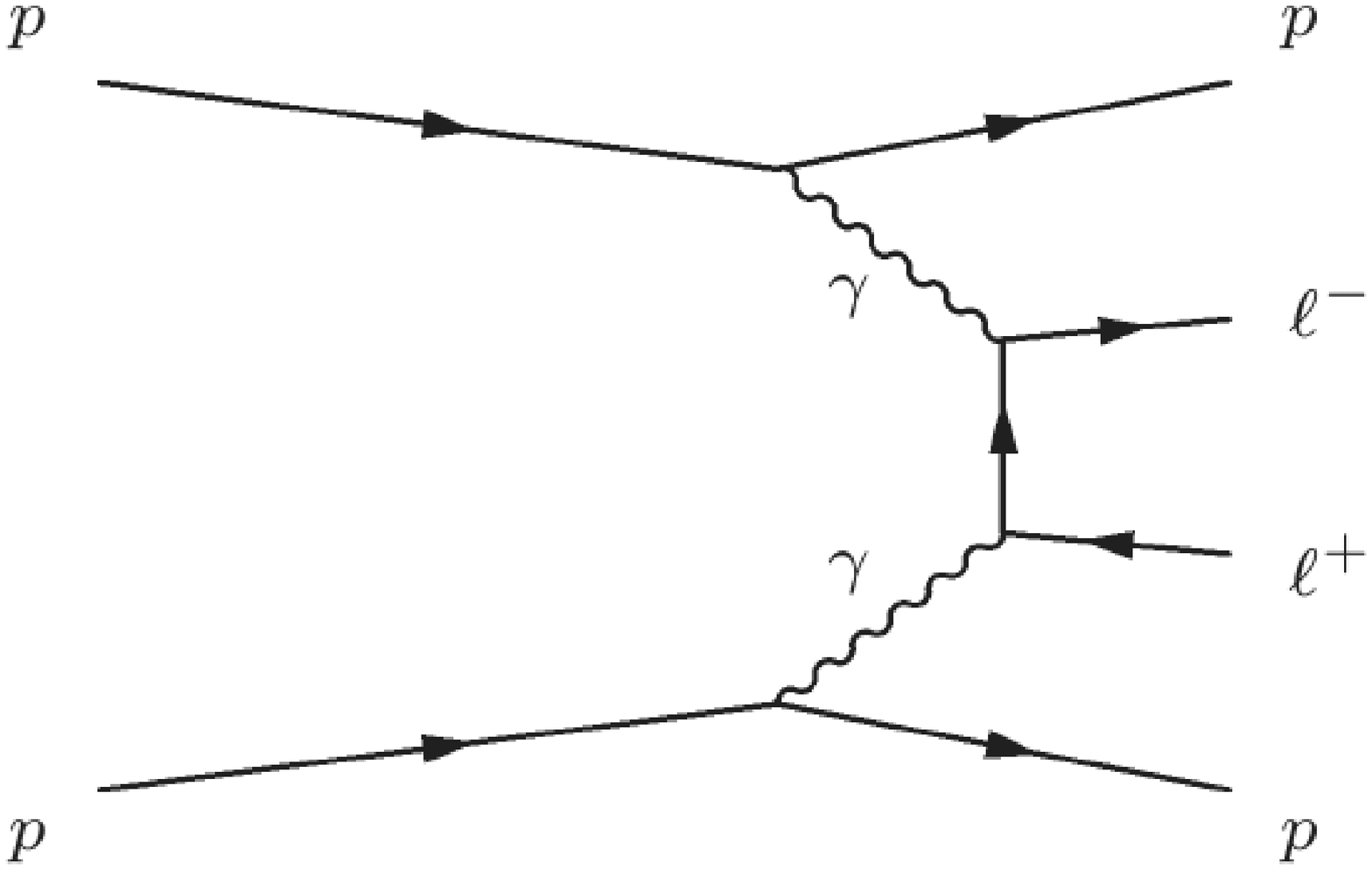,width=0.18\textwidth}
\psfig{figure=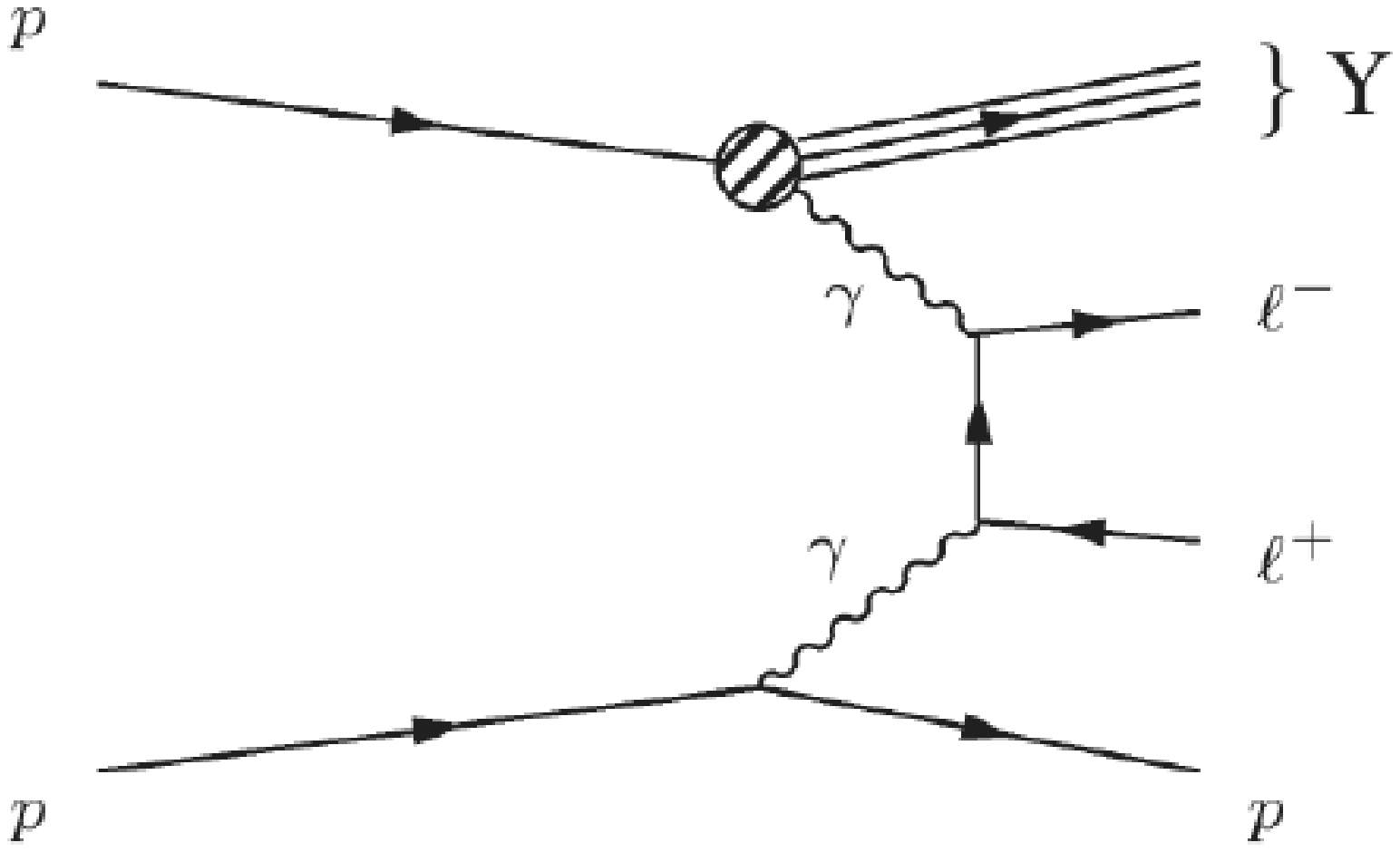,width=0.2\textwidth}
\psfig{figure=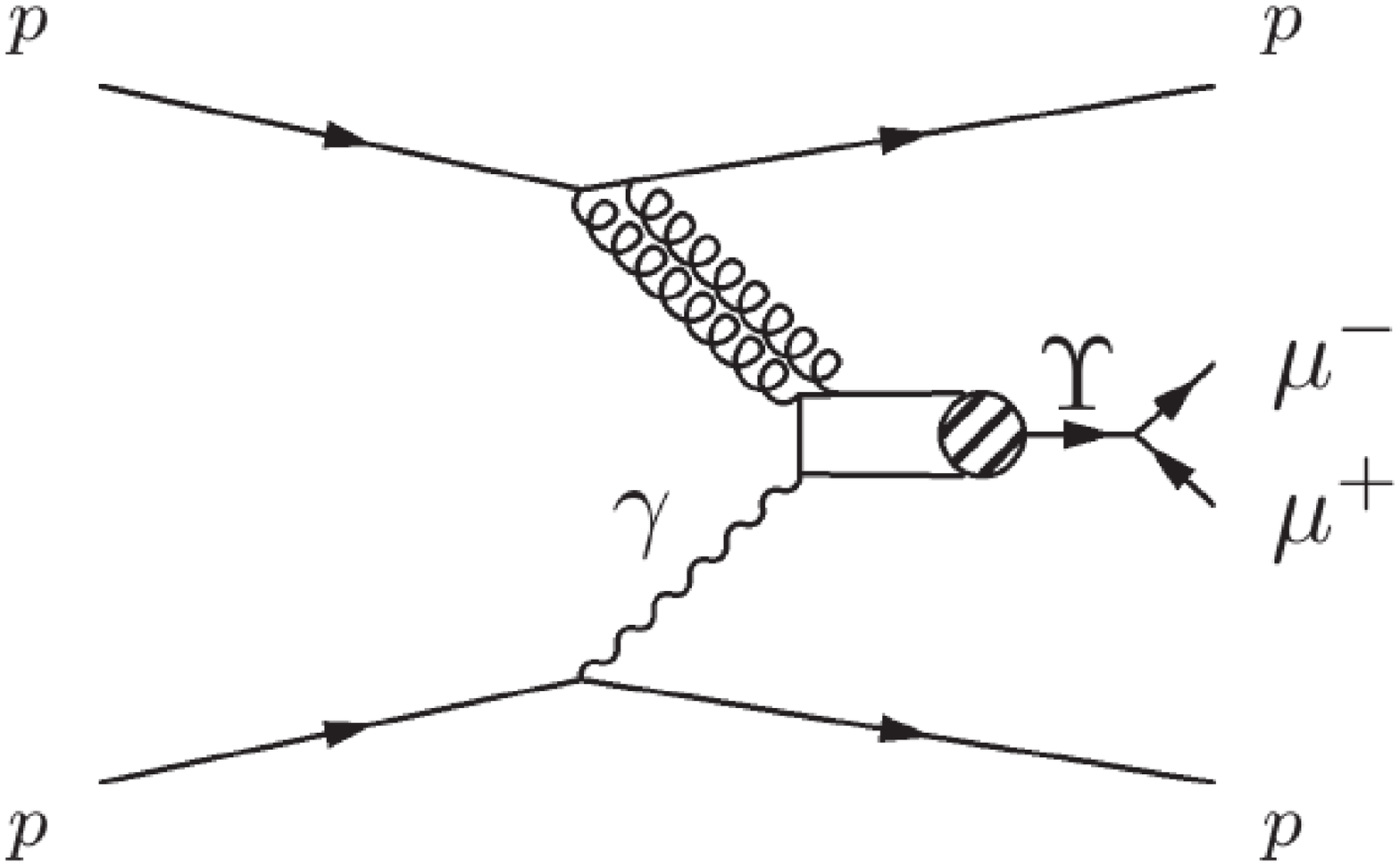,width=0.2\textwidth}
\caption{Exclusive production of muon pairs, for elastic (left) and singly-dissociative (middle) events, and from the decay of diffractive photoproduction of an $\Upsilon$ meson (right).
\label{fig:feynman}}
\end{figure}

\begin{figure}
\centering \psfig{figure=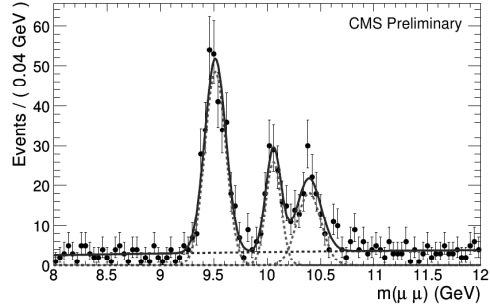,width=0.4\textwidth}
\caption{Dimuon invariant mass, for the selected sample, in the range $8 \leq M_{\mu^+\mu^-} \leq 12~\textrm{GeV}$. Dashed lines are results of a fit and full line their sum. The $\Upsilon(1S)$, $\Upsilon(2S)$ and $\Upsilon(3S)$ are visible above the $\gamma \gamma$ continuum.
\label{fig:upsilon}}
\end{figure}

In CMS, the production of exclusive dimuons ($pp \rightarrow \mu^+ \mu^- p p$) with two-photon interactions will provide from LHC start-up a sample of theoretically very well known events ($\Delta \sigma / \sigma < 1\%$), with a large statistics and a clean final state topology. After $100~\textrm{pb}^{-1}$, a selection based on balanced kinematics and exclusivity requirements yields a large number of elastic and singly-dissociative exclusive dimuons (Fig.~\ref{fig:feynman}):


\begin{equation}
\begin{split}
	&N_{elastic}(\gamma \gamma \rightarrow \mu^+ \mu^-) = 709 \pm 27 (stat) \pm 7 (th) \\
	&N_{inelastic}(\gamma \gamma \rightarrow \mu^+ \mu^-) = 636 \pm 25 (stat) \pm 121 (model) \\
\end{split} 
\end{equation}
The observed elastic events can be used for luminosity measurement, for a pile-up free environment, with a statistical uncertainty as low as $4\%$. A large amount ($\approx 2/3$) of inelastic events can be rejected using vetoes from CASTOR and ZDC calorimeters:


\begin{equation}
\begin{split}
	&N_{inelastic}^{w/ ~veto}(\gamma \gamma \rightarrow \mu^+ \mu^-) = 223 \pm 15 (stat) \pm 42 (model)
\end{split} 
\end{equation}
Exclusive dimuons can also serve for the alignment of forward detectors. Finally, the dimuon decay of exclusive production of $\Upsilon$ mesons ($pp \rightarrow \Upsilon pp \rightarrow \mu^+ \mu^- pp$, Fig.~\ref{fig:feynman}) also complies with the same selection criteria, allowing the observation of three resonances $(1S,2S,3S)$ above the $\gamma \gamma$ continuum (Fig.~\ref{fig:upsilon}). This allows the measurement of the process cross section and gives access to the gluon content of the proton (sensitivity to $t$ distribution slope). Moreover, such events provide alignment and calibration data at low $p_T$.
\begin{equation}
\begin{split}
	&N_{elastic}(\gamma p \rightarrow \Upsilon p \rightarrow \mu^+ \mu^- p) = 502 \pm 22 (stat) ^{+20}_{-150} (model)\\
\end{split} 
\end{equation}
The mean value of the effective $\gamma p$ center of mass energy in these events is $< W > = 2398 ~\textrm{GeV}$.

\section{Conclusions}
The ATLAS and CMS experiments will carry out a large physics program in forward physics, using a set of near-beam detectors (for tracking, calorimetry, timing and luminosity monitoring) along the LHC beamlines. This dedicated instrumentation (CMS/TOTEM : CASTOR, ZDC, FP420; ATLAS : LUCID, ZDC, ALFA ,RP220, FP420) provide the largest $\eta$ coverage ever in particle collider experiments. A large physics program will be covered, with amongst other things soft and hard diffractions, low-$x$ QCD, photon induced interactions, Higgs/SUSY/BSM, luminosity monitoring and measurements.

\section*{Acknowledgements}
The author would like to thank M. Grothe, P. Grafstr\"om and K. Piotrzkowski for useful discussions and comments.
Part of this work was supported by the Belgian Science Policy through the Interuniversity Attraction Pole P5/27 and by the European Commission through the Marie Curie Series of Events SCF.

\end{document}